\documentclass[prb,aps,twocolumn,noshowpacs,duperscriptaddresses]{revtex4-1}
\usepackage{graphicx}
\usepackage{dcolumn}
\usepackage{amsmath}
\usepackage{amssymb}
\usepackage{lmodern}
\usepackage{color}
\usepackage{longtable} 
\usepackage{units}
\usepackage{xcolor}

\usepackage{soulpos}
\definecolor{babyblue}{rgb}{0.54, 0.81, 0.94}

\newcommand{\kB}{k_{\rm B}}
\newcommand{\TC}{T_{\rm C}} 

\begin{document}
\widetext

\title{Modeling ultrafast all-optical switching in synthetic ferrimagnets}
\author{S.\ Gerlach$^1$}\email{stefan.gerlach@uni-konstanz.de}
\author{L.\ Oroszlany$^2$}
\author{D.\ Hinzke$^1$}
\author{S.\ Sievering$^1$}
\author{S.\ Wienholdt$^1$}
\author{L. Szunyogh$^3$}
\author{U.\ Nowak$^1$} 
\affiliation{$^1$ Fachbereich Physik, Universit\"{a}t Konstanz,
D-78457 Konstanz, Germany}
\affiliation{$^2$ Department of Physics of Complex Systems,
E\"{o}tv\"{o}s University, H-1117 Budapest,
P\'{a}zm\'{a}ny P\'{e}ter s\'{e}t\'{a}ny 1/A, Hungary}
\affiliation{$^3$MTA-BME Condensed Matter Research Group and
Department of Theoretical Physics,
Budapest University of Technology and Economics,
Budafoki \'ut 8., HU-1111 Budapest, Hungary}

\date{\today}

\begin{abstract}
Based on numerical simulations, we demonstrate thermally induced magnetic
switching in synthetic ferrimagnets composed of multilayers of rare-earth and
transition metals. Our findings show that deterministic magnetization reversal
occurs above a certain threshold temperature if the ratio of transition metal
atoms to rare-earth atoms is sufficiently large.
Surprisingly, the total thickness of the multilayer system has little effect
on the occurence of switching.
We further provide a simple argument to explain the temperature dependence of
the reversal process.
\end{abstract}

\maketitle

\section{Introduction} 
\label{sec:introduction}

The demonstration of helicity-dependent all-optical magnetization switching
\cite{stanciuPRL07,vahaplarPRL09} was one of the most surprising findings in
ultrafast magnetization dynamics. The experiments showed that magnetization
switching is possible solely triggered by a single laser pulse in the
sub-picosecond range avoiding any externally applied magnetic field.
These experiments were performed on GdFeCo, a rare-earth-based ferrimagnet
where the rare-earth (RE) sublattice is antiferromagnetically coupled to the
transition metal (TM) sublattice.
First attempts to describe these unexpected processes were based on the
assumption that the circularly polarized laser pulse induces a strong magnetic
field via the inverse Faraday effect which determines the direction of the
switching \cite{vahaplarPRL09, vahaplarPRB12}.
Over all, this process takes place on a time scale orders of magnitude shorter
than today's writing procedures in hard discs.
This calls for applications in magnetic data storage and alternative materials
including alloys \cite{alebrandPRB14}, heterostructures \cite{manginNatMar14}
and synthetic ferrimagnets \cite{tsemaINCOLLECTION15} are currently
investigated.

The discovery of thermally-induced all-optical switching using linearly
polarized light \cite{raduNature2011,ostlerNatComm12} cast a new light on
all-optical switching and called for more sophisticated models since this
switching works without any external or optically induced magnetic field,
which could define the magnetization direction during its recovery after the
ultrafast quenching. 

In simulations this switching was observed in an atomistic spin model developed
by Ostler \textit{et al.} \cite{ostlerPRB11,raduNature2011,ostlerNatComm12}.
An attempt to explain the occurrence of the transient ferromagnetic-like state
(TFMLS) was given by Mentink et al. who identified angular momentum transfer
driven by the inter-sublattice exchange as the crucial process
\cite{mentinkPRL12}.
Later on the thermally induced switching was more quantitatively described by
means of an orbital-resolved spin model, where the magnetic moments stemming
from $d$-electrons of the TM, the $d$-electrons of the RE and the $f$-electrons
of the RE are distinguished \cite{wienholdtPRB13}.
Here, it was shown that the initial laser excitation brings the sublattices into
a strong non-equilibrium after 1 ps.
This happens due to the different demagnetization times of the individual
sublattices \cite{frietschNatComm15}.
On that time scale, electron and phonon temperatures are nearly equilibrated
below $\TC$ again, but the Fe sublattice is already completely demagnetized
while the Gd sublattice remains still rather ordered.
The remagnetization dynamics of Fe taking place subsequently leads to a state
where the Gd spins and the Fe spins are aligned --- the transient
ferromagnetic-like state.
The  TFMLS arises naturally as a consequence of a redistribution of the energy
and angular momentum between the different sublattices due to the maximization
of entropy under the constraint of energy and angular momentum conservation.
These processes, which are driven via the precession term of the
Landau-Lifshitz-Gilbert equation, dominate on shorter times scales.
The following relaxation back to a ferrimagnetic equilibrium state is on a
longer time scale, where dissipative processes are responsible, and does not
necessarily lead to a switched state\cite{khorsandPRL12,alebrandPRB14}.
The details of this relaxation process depend on the material properties as
well as the experimental specifications and are still under investigation.

In the following, we will explore the possibility of thermally-induced switching
in synthetic ferrimagnets comprised of bilayers of Fe and Gd.
We use ab-initio methods to estimate spin model parameters for Fe-Gd bilayers.
The dynamic simulations of the spin model allows for an investigation of the
preconditions for thermally induced switching. 
We find that deterministic magnetization reversal occurs only above a certain
threshold temperature and in bilayers where the ratio of transition metal atoms
to rare-earth atoms is sufficiently large.
Finally, we find a simple explanation why the compensation temperature is so
important.

\section{Model} 
\label{sec:model}

Our aim is to model a synthetic ferrimagnet as a bilayer of two ferromagnets,
Fe and Gd, with a negative coupling between the two layers.
For that we consider an atomistic spin model where localized spins are arranged
on a simple-cubic lattice structure.
Spins experience exchange interactions with their nearest neighbors only and
dipole-dipole interaction is neglected.
The Heisenberg Hamiltonian of the system studied reads
\begin{equation}
{\cal H}=-\sum_{NN}J_{ij}{\mathbf S}_i{\mathbf S}_j-\sum_i d_z S^2_{i,z}.
\end{equation}
The first term represents the Heisenberg exchange energy, where the exchange
interaction is either between spins of the Fe layer, spins of the Gd layer, or,
across the interface, between Fe and Gd spins.
This term contains, therefore, three interactions, $J_{\rm Fe-Fe}$,
$J_{\rm Gd-Gd}$ and $J_{\rm Fe-Gd}$.
The second term represents a uniaxial anisotropy with anisotropy constant
$d_z$.
The lateral dimensions of the model are \mbox{150$\times$150} atoms with the
layers stacked along the $z$ axis and with periodic boundary conditions in
transverse directions. The thicknesses of the Fe and Gd layers are varied.

\textit{Ab-initio} calculations have been performed in terms of the fully
relativistic screened Korringa--Kohn--Rostoker (KKR) method, designed,
in particular, for layered systems and surfaces \cite{KKRBOOK}.
The LSDA parametrization from Ref.~\onlinecite{PZ-LSDA} was used.
The strong correlation of the localized $4f$-states of the Gd atoms was treated
within the framework of the LSDA+U approach \cite{Anisimov-1-LDAU} as
implemented within the KKR method \cite{Ebert-SPRKKR-LDAU}.
The calculations were carried out with the commonly used $U=6.7\mbox{ eV}$ and
$J=0.7\mbox{ eV}$ values of the Coulomb and exchange integrals
\cite{Anisimov-1-LDAU} and the double counting term derived in
Refs.~\onlinecite{Anisimov-double-counting-1,Anisimov-double-counting-2}
satisfying the atomic limit for the LSDA total energy.
The exchange constants have been obtained by means of the relativistic torque
method \cite{Udvardi-PhysRevB.68.104436}.

The geometry used in the \textit{ab-initio} calculations were based on a
heterostructure of six Fe layers between two semi-infinite bulk Gd regions.
For the hcp structure of Gd bulk we used the experimental $c/a$ ratio of 1.5904
and an optimized lattice constant $a=3.450$ \AA
\cite{Oroszlany-PhysRevLett.115.096402}.
The interlayer distance in the Fe region was chosen such that the volume per Fe
atom was identical to that in bcc Fe with the experimental lattice constant of
2.867 \AA.
The distance of the Fe planes to the nearest Gd planes was taken to be the
average of the Fe-Fe and Gd-Gd layer distances.
The interlayer exchange interactions obtained by the \textit{ab-initio}
procedure were then mapped to the simple cubic structure with nearest-neighbor
interactions used in the spin-dynamics simulations. 

The calculations outlined above result in exchange constants of the spin model
above with the ratios
$J_\text{Fe-Fe}:J_\text{Gd-Gd} :J_\text{Fe-Gd} =1:0.286:-0.388$
which we use in the subsequent dynamic simulations. 
The magnetic moments of Fe and Gd have the values of
$\mu_{\text{TM}} = 1.92 \mu_{\rm{B}}$ and
$\mu_{\text{RE}} = 7.63 \mu_{\rm B}$, refering to the values found for bulk
FeGd in Ref.~\onlinecite{wienholdtPRB13}.
These values are close to those obtained in our KKR LSDA+U calculations.
In addition, we used an anisotropy constant of $d_z = 0.2$ meV favoring
magnetization along the $z$ axis.

The dynamics of the system is governed by the stochastic Landau-Lifshitz-Gilbert
equation of motion,
\begin{eqnarray}  
           \frac{ (1+\alpha^2)\mu_i}{\gamma} {\dot{\bf{S}}}_i =
              -   \mathbf{S}_i  \times  \left[ \mathbf{H}_i 
              + \alpha \;   \left( \mathbf{S}_i  \times \mathbf{H}_i \right)
		\right],
\end{eqnarray}
with the gyromagnetic ratio $\gamma$ and a dimensionless Gilbert damping
constant $\alpha$ that describes the coupling to the heat-bath.
In our simulations the damping constant is set to $\alpha = 0.02$
\cite{wienholdtPRB13, schlickeiserPRB12}.
Thermal fluctuations are included as an additional white noise term $\boldsymbol{\zeta}_i$
in the internal fields
$\mathbf{H}_i= - \frac{\partial {\cal H}}{\partial \mathbf{S}_i}
+ \boldsymbol{\zeta}_i(t)$
with  
\begin{equation}
\langle{\boldsymbol \zeta}_i(t) \rangle = 0, \quad 
 \langle\zeta_{i\eta}(0) \zeta_{j\theta}(t) \rangle = 
\frac{2 \kB T \alpha \mu_i}{\gamma} \delta_{ij}\delta_{\eta\theta} \delta(t), 
\end{equation}
where $i,j$ denotes lattice sites and $\eta,\theta$ are Cartesian components.
All algorithms we use are described in detail in Ref.~\onlinecite{nowakBOOK07}.

\begin{figure}
\includegraphics[width=\columnwidth]{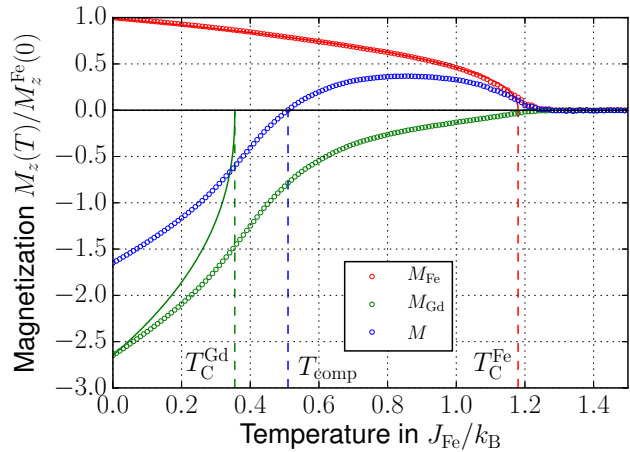}
\caption{(Color online) Temperature dependent equilibrium magnetization for a
bilayer with 3 monolayers of Fe and 2 monolayers of Gd.
The magnetization of the element with the stronger intra-layer exchange
interaction, in this case Fe, closely follows the shape of a ferromagnet with
its Curie temperature $T_\text{C}^\text{Fe}$ close to the Curie temperature of
the coupled system.
The element with the weaker intra-layer exchange interaction, in this case Gd,
exhibits magnetic ordering above its own bulk Curie temperature,
$T_\text{C}^\text{Gd}$, due to the interaction with the other sublattice. 
The chosen ratio of layer thicknesses leads to a larger magnetic moment of the
Gd layer at low temperatures and, consequently, to a magnetic compensation point
at a temperature $T_\text{comp}$ which is slightly above the bulk
Curie temperature of the Gd.}
\label{f:mag}
\end{figure}

In Fig.~\ref{f:mag} we present the equilibrium properties of a bilayer
consisting of 3 monolayers of Fe and 2 monolayers of Gd.
This bilayer behaves as a synthetic ferrimagnet with a magnetic compensation
point $T_{\rm comp}$ at a temperature of about 40\% of the Curie temperature
$T_c$. 

It is well-known that laser-induced demagnetization in transition metals is
several times faster than in rare-earth elements
\cite{vaterlausPRL91,beaurepairePRL96}.
Several approaches have been proposed to explain this behavior, including
electron-phonon scattering processes of the Elliott-Yafet type
\cite{koopmansNATMAT10} and intra-atomic energy transfer within the electronic
subsystem \cite{wienholdtPRB13}.
However, based on the different demagnetization time scales, we may assume that
heating a multilayer system with incident laser light can lead to a situation
where the TM layer is completely demagnetized while the RE layers still retains
a substantial net magnetization.
We use this fact in the following and do not calculate the action of the laser
pulse on the spin system explicitly.
Instead we focus on the relaxation of the magnetization at constant temperature
starting our simulations with a spin configuration where the Fe sublattice is
completely demagnetized (spins are randomly oriented) while we vary the degree
of magnetization of the Gd layer.
This initial Gd magnetization and the temperature will turn out to be crucial
quantities for the understanding of thermally triggered switching. 

\section{Results} 
\label{sec:results}

\begin{figure}
\includegraphics[width=.9\columnwidth]{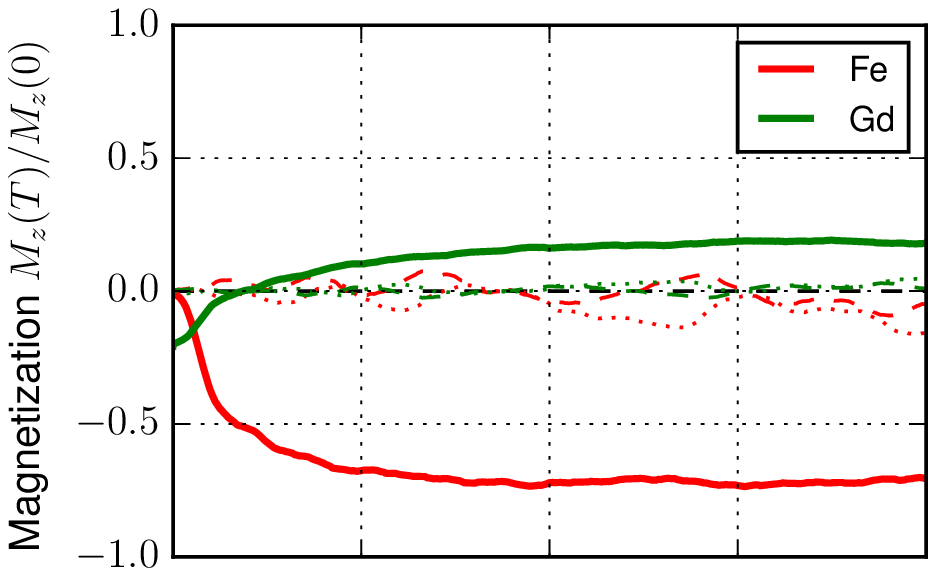}\\[-4mm]
\includegraphics[width=.9\columnwidth]{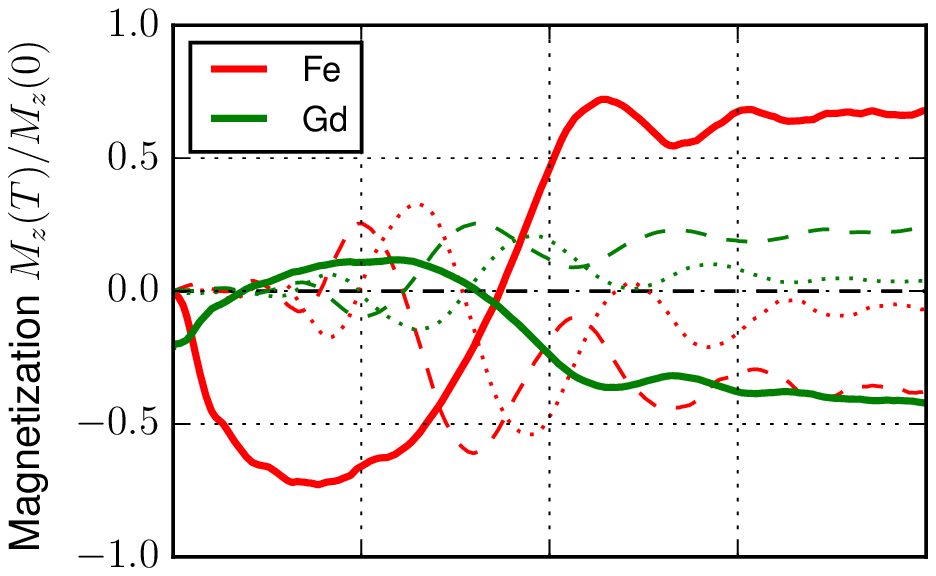}\\[-4mm]
\includegraphics[width=.9\columnwidth]{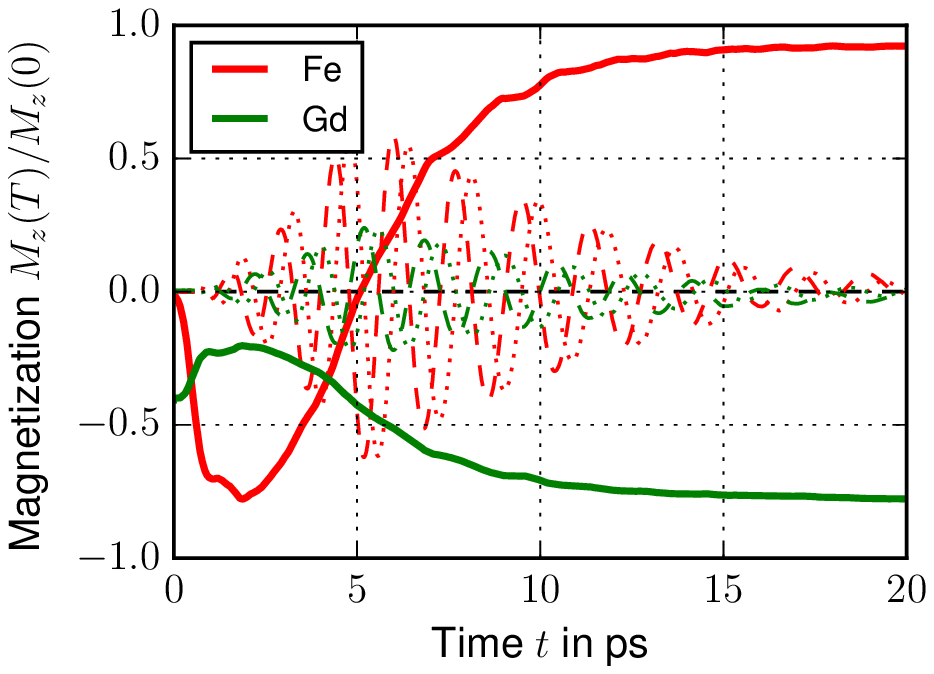}\\[-4mm]
\caption{ (Color online) The three possible time relaxation scenarios for the
sublattice magnetizations of a RE-TM layered system.
(top) Magnetization switches with respect to the initial configuration,
(center) both sublattice magnetizations change their signs twice, ending up back
in the initial state, (bottom) no-switching, i.e. the rare-earth layer
magnetization does not change sign.
Also shown are the transverse components (dotted and dashed lines).
For the case of back-switching and no-switching these components are of the
order of the magnitude of the magnetization.}
\label{f:var_mag}
\end{figure}

\begin{figure}
\includegraphics[width=\columnwidth]{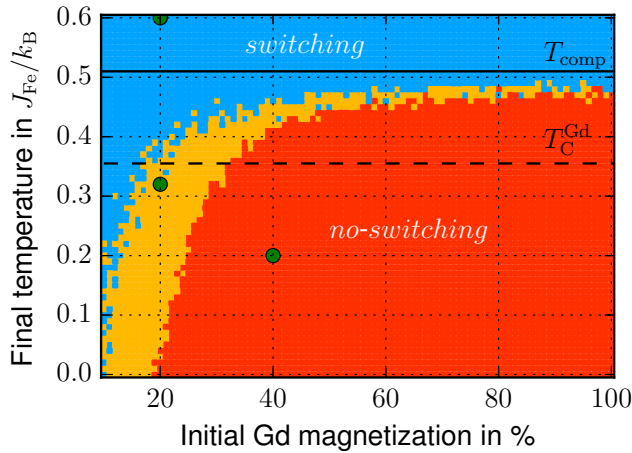}
\caption{ (Color online) Switching in a 3:2 layer sample after 40 ps.
Color coding: {\color{cyan} $\blacksquare$} Magnetization switches,
{\color{red} $\blacksquare$} magnetization does not switch and returns to the
initial state, {\color{orange} $\blacksquare$} both magnetizations change their
sign twice, ending up in the initial state (back-switching).
The compensation temperature $T_\text{comp}$ is indicated by a solid black line,
the Curie temperature $T_\text{C}^\text{Gd}$ of the isolated Gd layer by a black
dashed line.
The three points indicate the chosen values for the scenarios in
Fig.~\ref{f:var_mag}.}
\label{f:switching}
\end{figure}

For a fixed layer configuration, we treat the initial RE magnetization remaining
after the laser excitation and the temperature $T$ as the relevant parameters
which determine the magnetization dynamics triggered by the laser pulse.
In Fig.~\ref{f:var_mag} three different possible scenarios are shown, switching
(top), switching followed by switching back to the initial state (center), and
no-switching (bottom).
The chosen values for $T$ and Gd magnetization are indicated in
Fig.~\ref{f:switching}.
The switching scenario corresponds to the work of Radu et al.
\cite{raduNature2011}, while the back-switching was measured by Khorsand et al.
\cite{Khorsand-PRL.110.107205}. 
Note, that in all three cases the magnetization of the transition metal starts
towards negative values (while the original sign before demagnetization by the
effect of the laser heating would have been positive) so that a TFMLS is
obtained.
Note also, that the transverse components of the magnetization are usually not
small --- apart from the case of switching --- which indicates a linear
mechanism for the case of successful switching but a more precessional process
for the case of no-switching and back-switching.

A systematic variation of the two parameters, temperature and initial Gd
magnetization, allows for the construction of a switching diagram as shown in
Fig.~\ref{f:switching}.
It is ternary in the way that it provides information about the relaxation
process, with the three scenarios above (switching, back-switching, or
no-switching) as shown in Fig.~\ref{f:var_mag}.
The diagram is constructed by taking into account the magnetization dynamics of
the two species involved for a single run during a time interval of 40 ps.
We identify three distinct regions, a connected no-switching region
({\color{red} $\blacksquare$}), a connected switching region
({\color{cyan} $\blacksquare$}), and a back-switching region along the boundary
of the other two regions ({\color{orange} $\blacksquare$}).

For an interpretation of this diagram we first note that for very low values of
initial RE magnetization one would expect the system to randomly pick one of
the two possible equilibrium configurations, either switched or not.
This sort of statistical behavior is indeed evident from the isolated data
points at the very left hand side of the diagram.
Furthermore one would expect that high values of $M_{\rm{RE}}$ would prevent
the system from switching at low temperatures because the level of order in the
RE layer is too high to become demagnetized.
This is indeed what we find from Fig.~\ref{f:switching}.
The no-switching region is found in the bottom right corner, corresponding to
low temperature and high $M_{\rm RE}$.

So far, the switching diagram meets our intuitive expectations.
It is less obvious, however, why high temperatures (above $T_\text{comp}$) lead
to switching regardless of how strongly the RE layers are demagnetized by the
heat pulse.
Qualitatively, this can be understood keeping in mind the linear switching
mechanism, which avoids transverse magnetization components (see
Fig.~\ref{f:var_mag} and Refs.~\cite{vahaplarPRL09, kazantsevaEPL09}).
Linear switching needs a high degree of spin disorder in the system and,
consequently elevated temperatures. 

More quantitatively, the role of temperature can be understood via the
equilibrium layer magnetizations as shown in Fig.~\ref{f:mag}, since those mark
the final values for the relaxation process.
Let us assume that following the excitation of the laser pulse the Fe layer is
completely demagnetized while the Gd is only demagnetized by 50\%.
The bilayer is far from equilibrium and a relaxation process will set in of
which the details depend on the temperature.
Each layer will relax towards its individual equilibrium value.
We can identify three important temperature ranges:
\begin{itemize}
\item $T < T_{\rm C}^{\rm{RE}}$: Both layers tend to increase their net
magnetization magnitudes towards higher values.
\item $T_{\rm C}^{\rm{RE}} < T < T_{\rm C}^{\rm{TM}}$: To reach equilibrium,
the magnitude of the TM magnetization must still increase, while that of the RE
must decrease.
\item $T > T_{\rm C}^{\rm{TM}}$: The system will completely demagnetize.
\end{itemize}

For most TM-RE layer ratio $T_\text{comp}$ (if it exists) is higher than
$T_{\rm C}^{\rm{RE}}$ and only the temperature range between
$T_{\rm C}^{\rm{RE}}$ and $T_{\rm C}^{\rm{TM}}$ supports the dynamics necessary
for switching --- decreasing Gd magnetization and increasing Fe magnetization.
Consequently, the temperature must be at least above the critical temperature
of bulk Gd.
Above the compensation temperature switching is always possible.
That means, if $T_\text{comp}$ is very low, switching can also be done below
$T_{\rm C}^{\rm{RE}}$.

The next question is, why the Fe layer magnetization starts recovering towards
negative magnetization which results in a TFMLS.
This is a consequence of angular momentum conservation, as was already pointed
out by several authors \cite{mentinkPRL12, wienholdtPRB13}.
While the Gd still demagnetizes (since the temperature is above the bulk
critical temperature of Gd), the dynamics of the Fe layer magnetization must
change into the other direction keeping the angular momentum constant.
The argument also explains that one needs a certain initial Gd magnetization to
start with.
Without initial Gd magnetization there is no angular momentum reservoir to drag
the Fe magnetization towards negative values. 

\begin{figure}
\includegraphics[width=\columnwidth]{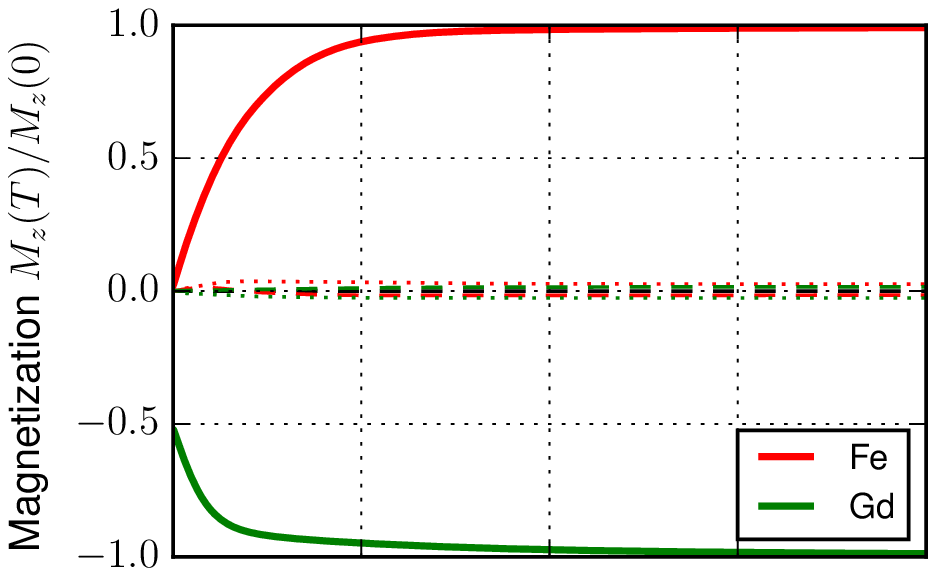}\\[-4mm]
\includegraphics[width=\columnwidth]{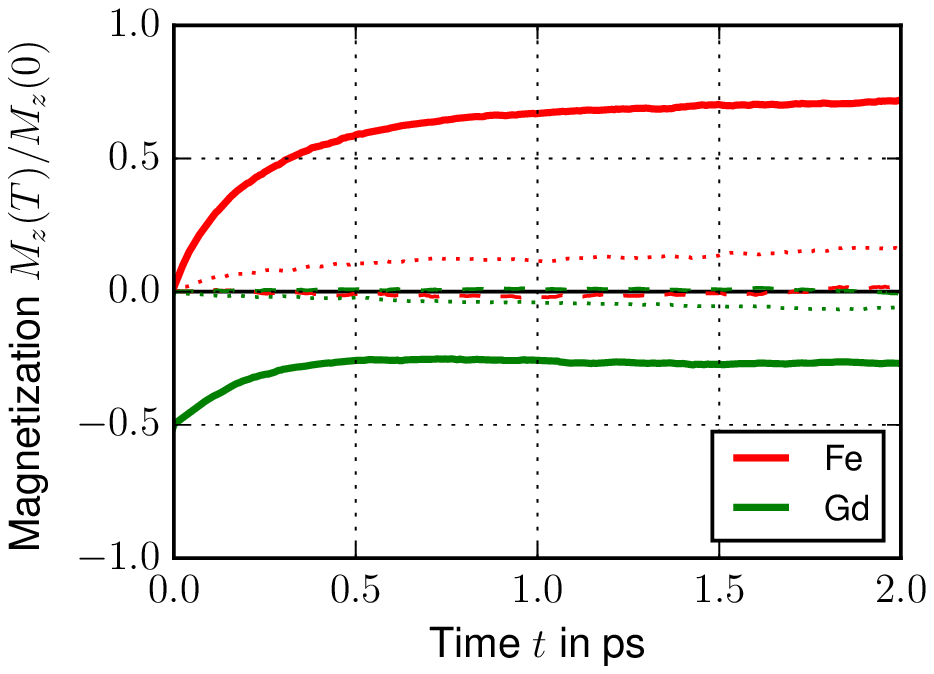}\\[-4mm]
\caption{(Color online) Strong dissipation ($\alpha=1$) in a 3:2 layer sample,
implying no conservation of angular momentum.
The relaxation does not lead to a TFMLS, neither at low temperature where the Gd
magnetization increases nor at high temperature where it decreases.
(top: $T<T_\text{C}^\text{Gd}$, bottom: 
$T_\text{C}^\text{Gd} < T < T_\text{C}^\text{Fe}$).
Also shown are the transverse components (dotted and dashed lines) which remain
small for both cases.}
\label{f:strong-diss}
\end{figure}

Angular momentum conservation is not strictly fulfilled in the spin system.
The relaxation part of the equation of motion breaks this conservation on time
scales which are determined by the value of the damping constant $\alpha$.
For low values of $\alpha$, the precessional part of the equation of motion is
much larger leading to dynamics which keeps total energy and total angular
momentum conserved on shorter time scales.
This is different when considering larger values of $\alpha$.
For comparison, we investigate in Fig.~\ref{f:strong-diss} the regime of strong
dissipation ($\alpha = 1$).
Here, the time scales of precessional dynamics is of the same order as the time
scale of relaxation and dissipative effects counteract the conservation of
angular momentum in the system.
Only if the damping constant $\alpha$ is sufficiently small, angular momentum
is almost conserved on short time scales, along with the total energy, leading
to a TFMLS as seen in Fig.~\ref{f:var_mag}.
The figure also illustrates that --- depending on the temperature --- the Gd
magnetization might relax either towards higher or lower equilibrium values. 
The transition from dissipationless dynamics to the regime where damping effects
dominate the dynamics has previously been investigated\cite{wienholdtPRB13}.

\begin{figure}
\includegraphics[width=\columnwidth]{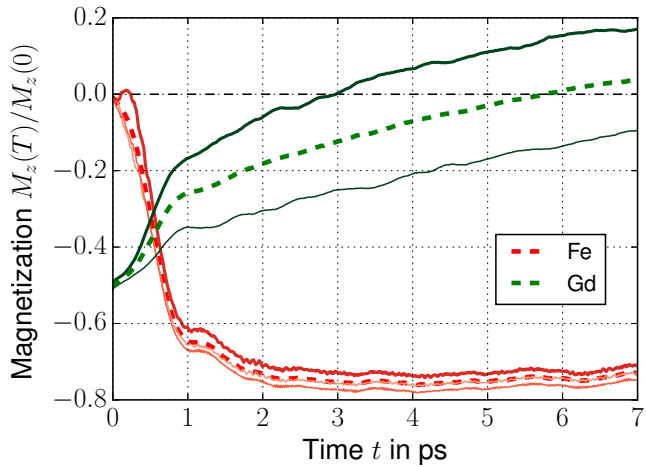}
\caption{(Color online) Layer resolved magnetization dynamics for switching in a
3:2 layer sample.
The Gd layer is demagnetized to 50\% of its zero temperature value.
The bold lines corresponds to the layers at the Gd-Fe interface and the dashed
lines shows the average value for Gd or Fe.
The temperature is $T = 0.5 J_\text{Fe}/k_\text{B}$.}
\label{f:layer_resolved}
\end{figure}

In the following we turn to the peculiarities of the layered system.
The layer resolved magnetization in Fig.~\ref{f:layer_resolved} shows the
importance of the interface layers for switching.
For low damping, the angular momentum conservation leads to a relaxation
dynamics with an exchange of angular momentum between the still demagnetizing
Gd layer and the Fe layer, leading to a negative Fe magnetization and,
consequently, to a TFMLS.
Because of the antiferromagnetic coupling along the Fe-Gd interface, the Fe
interface monolayer lags behind the other layers.
After some ps, however, the Fe magnetization has reached its new, negative
equilibrium value, pushing the Gd via the negative interface coupling towards
positive values.
Here, the dynamics is quicker at the interface as for the other Gd layer which
is lagging behind. 

We also simulated other layer thicknesses and ratio.
While Fe-Gd bilayer with 20-40\% Gd (for example 4:1, 3:1 or 3:2 layer) turned
out to have the correct thickness ratio for switching (and having a
magnetization compensation temperature) we found that the over-all thickness is
less relevant.
Successful switching can also be seen in much bigger samples (although on longer
time scales), for instance in a 30:20 Fe-Gd layer.
The antiferromagnetic coupling of the interface layers finally leads to a
switching of all layers when the temperature of the heat bath exceeds
$T_\text{comp}$ (which is almost the same as $T_\text{C}^\text{Gd}$ for big
samples).
Changing the ratio of Fe and Gd layer does not change the switching behavior as
long as the sample maintains a magnetization compensation temperature.
Only the temperature range for switching increases with decreasing percentage
of Gd.

\section{Summary}   
We explored thermally induced magnetic switching in synthetic ferrimagnets
composed of a bilayer of rare-earth and transition metal on the basis of spin
model simulations where the model parameters were calculated from first
principles.
Varying the temperature and the degree of initial rare-earth magnetization
directly after the laser pulse one may find either, back-switching, or
no-switching.
Deterministic magnetization reversal occurs above a certain threshold
temperature which is above the bulk Curie temperature of the rare-earth since
only then the magnetization of the rare-earth sublattice relaxes towards lower
magnitude.
The optimal ratio of transition metal atoms to rare-earth atoms for successful
switching has 20-40 \% Gd layer while the total thickness of the multilayer
system only affects the time scale of switching. 

\begin{acknowledgments} 
This work has been funded by the Center for Applied Photonics at the University
of Konstanz, the {\em Deutsche Forschungsgemeinschaft} and by the Hungarian
National Scientific Research Fund (NKFIH) under project Nos. K115575 and
K108676.
O. L. acknowledges support from the Janos Bolyai Scholarship of the Hungarian
Academy of Sciences.
\end{acknowledgments}

\bibliographystyle{apsrev4-1}
\bibliography{Cite}

\end{document}